\documentstyle[12pt]{article}
\newcommand{\be}{\begin{equation}}
\newcommand{\ee}{\end{equation}}
\newcommand{\bea}{\begin{eqnarray}}
\newcommand{\eea}{\end{eqnarray}}
\newcommand{\eV}{\,{\rm eV}}

\newcommand{\GeV}{\,{\rm GeV}}

\newcommand{\ns}{\normalsize}

\begin{document}
\begin{titlepage}

\title{\large\bf Supersymmetric Singlet Majorons and Cosmology}
\author{ E.~J.~Chun\\[0.2cm]
        {\ns International Center for Theoretical Physics}\\
        {\ns P.~O.~Box 586, 34100 Trieste, Italy}\\[2mm]
        {\ns and}\\[2mm]
        H.~B.~Kim\\[0.2cm]
        {\ns Center for Theoretical Physics and Department of Physics}\\
        {\ns Seoul National University}\\
        {\ns Seoul 151-742, Korea}\\[2mm]
        {\ns and}\\[2mm]
        A.~Lukas\\[0.2cm]
        {\ns Physik Department}\\
        {\ns Technische Universit\"at M\"unchen}\\
        {\ns D-85748 Garching, Germany}\\
        {\ns and}\\[0.1cm]
        {\ns Max-Planck-Institut f\"ur Physik}\\
        {\ns Werner-Heisenberg-Institut}\\
        {\ns P.~O.~Box 40 12 12, Munich, Germany}}

\date{}

\maketitle

\begin{abstract}
We examine cosmological constraints on the lepton number breaking scale
in supersymmetric singlet majoron models. Special attention is drawn to
the model dependence arising from the particular choice of a certain
majoron extension and a cosmological scenario. We find that the bounds
on the symmetry breaking scale can vary substantially. Large values of this
scale can be allowed if the decoupling temperature of smajoron and majorino
exceeds the reheating temperature of inflation. In the opposite case an upper
bound depending on the majoron model can be obtained which, however, is
unlikely to be much larger than $10^{10}$ GeV.
\end{abstract}

\vskip -23.5cm
\hfill \parbox{5cm}{IC/94/40 \\  SNUTP 94-15 \\ TUM - TH - 164/94 \\
                    hep-ph/9403217 \\ February 1994}

\thispagestyle{empty}
\end{titlepage}
\clearpage

\setcounter{page}{1}
\baselineskip 6.5mm

Singlet majoron models~\cite{cmp} provide the simplest extension of the
standard model with spontaneously broken lepton number and neutrino masses
generated via the celebrated see-saw mechanism. Therefore these models are
of particular interest for cosmology.
The well-known cosmolgical constraint on the conventional singlet majoron model
comes from the late-decay of heavy neutrinos into majorons ($J$) and
lighter neutrinos. In the model with one singlet $S$
the amplitude for this decay  has a strong suppression~\cite{sv}.
Then the standard cosmological constraint derived from the closure density
of the universe~\cite{dkt} bounds the right handed neutrino (RHN) mass to be
$< 10^3$ GeV taking the largest possible neutrino mass of 30 MeV and a mixing
angle of order $10^{-4}$. For generic models with an arbitrary number of
singlets the bound can be shifted to $\sim 10^9$ GeV. Clearly, for
decreasing neutrino masses these bounds become weaker and for masses below
$\sim 40$ eV no information about the RHN scale can be obtained at all.

In supersymmetric models, cosmological considerations provide independent
informations due to the presence of the superpartners of the majoron
(the  smajoron $\sigma$ and the majorino $\psi$) as discussed first by
Mohapatra and Zhang~\cite{mz}. Because their coupling is surpressed by
the lepton number symmetry breaking scale $V_L$ they can be expected to
decouple when relativistic. In comparison to supersymmetric axion
models~\cite{rtw} the partners of the majoron tend to couple much weaker
to the standard particles due to the lack of an anomaly and the smallness of
the
neutrino masses. One might therefore ask if the constraints
on $V_L$ from baryogenesis and overclosure will be more restrictive than
corresponding constraints on axion models. We consider the range
{}from $\sim 10^{10}$ GeV to $10^{12}$ GeV as particulary attractive for
$V_L$ since several independent hints from the MSW solution of the solar
neutrino problem, the hidden sector SUSY breaking and axion physics point to
such scales. A question of particular interest might therefore be to what
extent such large values of $V_L$ are compatible with cosmological
restrictions on SUSY majoron models.

In the below we examine this matter taking two kinds
of model dependences
into account. On the one hand we discuss the freedom in the field theoretical
parameters relevant for cosmology like masses and couplings of the smajoron and
the majorino. On the other hand we analyze the influence of the cosmological
scenario, namely the influence of inflation which for large $V_L$ may be
able to wash out the relic densities of smajoron and majorino.

\bigskip

We start by briefly introducing the class of models we are considering here.
Their superpotentials which come in addition to that of the MSSM
can be written as
\be
  W = h_{ij} L_i H N_j + f_{ija}N_i N_j S_a + W(S_a) \label{superpot}
\ee
with three families $N_i$ of RHNs and  gauge singlet fields $S_a$.
For most of our discussions soft breaking terms of the standard type
\be
 V_{soft} = m_{3/2}^2\sum_a |\phi_a|^2+m_{3/2}\left( \sum_a\phi_a
            {\partial W \over \partial\phi_a}+(A-3)W + {\rm h.~c.}\right)
 \label{softterms}
\ee
are assumed. We have in mind that supergravity is broken in the hidden
sector.  The whole potential is invariant under $U_L(1)$ with $Q_L(L)=1$,
$Q_L(N)=-1$ and $Q_L(S_a)=q_a$,
which is spontaneously broken by the VEVs $v_a$ of
the gauge singlet fields $S_a$. Apart from being larger than $m_{3/2}$ the
$U_L(1)$ symmetry breaking scale $V_L=(\sum |v_a|^2 q_a^2)^{1/2}$ is taken
as a free parameter. The freedom for model building consists in the number
of singlet fields $S_a$ and the choice of their potential $W(S_a)$.
In this paper we assume that R-parity is unbroken.

\bigskip

For cosmological considerations we extract now the model-dependence of
masses and lifetimes of smajoron and majorino.
Let us first examine  the interactions of the majoron supermultiplet
$\Phi = ( (\sigma + iJ)/\sqrt{2}, \psi)$. The interaction with the standard
particles is given by the usual form
\be \label{masu}
 W = h L H N + {M\over 2} N N + {f\over 2} N N \Phi
\ee
with the generation indices suppressed. We also need to know higher-dimensional
effective interactions. The one most important for our purpose comes from a
one loop diagram built up from the interactions $LHN$ and $\Phi NN$.
It leads to the effective D=5 operator $(\Phi +\bar{\Phi})L\bar{L}$.
Putting the fields on shell the diagram vanishes so that there is no
decay mode of $\Phi$ with a $1/V_L$ surpression. However, attaching
a gauge field $V$ gives the coupling
\be \label{d5}
 \frac{gh^2}{16\pi^2 V_L}(\Phi+ \bar{\Phi})L\bar{L}V \, .
\ee
Here $h$ denotes the dominant one of Yukawa couplings $h_{ij}$ in
eq.~(\ref{superpot}). This diagram can keep the majoron supermultiplet
in equilibrium after the phase transition.

To describe the interactions among the components of $\Phi$ which
generally depend on the form of $W(S_a)$,
the nonlinear approach is easier to handle
since the majoron and the majorino are completely rotated
away from the superpotential as well as from the soft terms.
Only the K\"ahler part needs to be considered.
Starting with the minimal K\"ahlerian $K=\sum_a S_a\bar{S_a}$
and writing $S_a = v_a \exp (q_a\Phi / V_L)$,
we arrive at the following Lagrangian in component fields~:
\bea
 \cal{L}&=&(1+\frac{\sqrt{2}x}{V_L}\sigma )(\frac{1}{2}\partial_\mu
 \sigma\partial_\mu \sigma + \frac{1}{2}\partial_\mu J \partial_\mu J +
 i\bar{\psi}\sigma_\mu \partial^\mu\psi) \nonumber\\
 &&-\frac{x}{V_L}(\partial^\mu J)\bar{\psi}\sigma_\mu \psi +
  (4\; {\rm fermion}\; {\rm terms}) + O(1/v^2) \\
 & {\rm with} &  x = \sum_a \frac{|v_a|^2}{V_L^2}q_a^3 \;.
\eea
The model dependence is now condensed in the value of $x$ which influences
the decay rates for  $\sigma\rightarrow 2 J$ and $\sigma\rightarrow 2\psi$.

The value of $x$ can be of order one or zero at tree level.
Even in the second case a nonzero value can be generated
by radiative corrections. To discuss this, let us take the simplest model.
It is specified by the superpotential
\be
  W = h_{ij} L_{\alpha i} H_\alpha N_j
      + f_{ij}N_i N_j S + \lambda (SS'-\mu^2)Y
\ee
which results in the following soft terms for the gauge singlet fields~:
\be
 V_{soft} = \sum_{\phi = n,s,s'}m_\phi^2 |\phi|^2
           + m_{3/2}\left[ A_n fnns+A_s\lambda ss'y-A_y\lambda\mu^2 y
           +{\rm h.~c.} \right] \; .
\ee
The boundary condition to be fulfilled at $M_{Pl}$ is $m_\phi = m_{3/2}$,
$A_n=A_s=A$ and $A_y=A-2$. The value of $x$ for the above potential is
$x=(v^2-{v'}^2)/(v^2+{v'}^2)$. Minimization leads to
\be
 x=\frac{m_{s'}^2-m_s^2}{2\lambda^2 y^2+m_s^2+m_{s'}^2}
  \sim \frac{m_{s'}^2-m_s^2}{4m_{3/2}^2}\; ,
\ee
with $y=(A_y-A_s)m_{3/2}/2\lambda$.
Obviously degenerate soft masses result in $x=0$ which is not surprising since
in that case a permutation symmetry $s\leftrightarrow s'$ exists.
The only way to break
this symmetry is to take $m_s\ne m_{s'}$ which can be achieved in two
ways~: One can assume nonuniveral mass terms at $M_{Pl}$ thereby
departuring from the standard soft terms~(\ref{softterms}) or one can
consider renormalization group effects. The first option clearly allows to
produce any value of $x$. So let us discuss the
implications of RGE starting with universal boundary masses.
Due to the asymmetric coupling of $s$ and $s'$ to the RHNs their masses
renormalize differently and one finds~\cite{fal}
\bea \label{x}
 x&\sim& \frac{f^2}{64\pi^2}(1+A^2)\ln\left(\frac{M_{Pl}}{V_L}\right)
                                \nonumber \\
  &=&f^2(1+A^2)\left( 2.2\times 10^{-2}+\frac{1}{64\pi^2}\ln\left(
     \frac{10^{12}{\rm GeV}}{V_L}\right)\right)\; .
\eea
So it appears to be difficult to get $x>0.1$.
The lifetime of the smajoron decay $\sigma\rightarrow 2J$ is given by
\be \label{ts}
 \tau= 2\times 10^{-5}
         \left(\frac{V_L}{x10^{12}{\rm GeV}}\right)\left(\frac{10^2{\rm GeV}}
      {m_\sigma}\right)^3 {\rm sec} \; .
\ee

In another model discussed in ref.~\cite{ckn} only one change appears.
The value of $y$ is negligible, and so the value of $x$ roughly doubles.

\bigskip

The majoron has mass zero forgetting about possible $U_L(1)$-breaking gravity
effects. Considering this effect appearing in the D=5 operators it was found
that the cosmological mass density constraint gives a strong bound on the
lepton number breaking scale~: $V_L < 10$ TeV \cite{abms}.  In this paper we
neglect this effect.   The smajoron cannot escape receiving a
mass $m_\sigma \sim O(m_{3/2})$ from soft terms.
The majorino tree level mass is model-dependent and can
take values $m_{\psi ,tree}\sim O((m_{3/2}/V_L)^k m_{3/2})$ with
$k=0$ or $1$ typically. A small tree level majorino mass ($k>0$) is
e.~g. realized in the model of
ref.~\cite{ckn}\footnote{This model can provide a light majorino
with the standard universal soft terms.
But it has another zero mode than the majoron,
which may cause additional cosmological impacts~\cite{cckl}.}
or in the above model for nonstandard soft terms satisfying
$A_s=A_y$ at $M_{Pl}$. However, in analogy to the heavy quark axion \cite{my},
it receives a {\it model-independent} one-loop correction
$m_{\psi ,loop}=(1/16\pi^2)f^2Am_{3/2}$, where $f$ is the dominant one of
the couplings $f_{ija}$. Additionally, there can arise another
radiative mass due to the RG evolution of the trilinear soft couplings in
such models. The induced mass depends on $f$ and the completely
unrestricted couplings in the $W(S_a)$-sector~\cite{cckl} and is therefore
quit model-dependent. In any case the majorino mass can take values far
below $m_{3/2}$. To illustrate the cosmological implications
we will use the above model-independent value $m_{\psi, loop}$ and
examine the cosmological effects depending on the value of $f$.

\bigskip

When the temperature of the universe falls below the scale $V_L$,
smajoron and majorino can be kept in equilibrium via effective interactions
with light fields.  Considering the D=5 effective interaction in
eq.~(\ref{d5})
smajoron and majorino get out of equilibrium below the temperature
\be \label{tl}
 T_L =  6.2 \times 10^{17} f_1^{-2}\sqrt{g_*/100}
           \left( 10 \eV \over m_\nu \right)^2 \GeV  \,.
\ee
Hereafter we introduce the notation $f_n = 10^n f$.
The actual decoupling temperature is the lower one between $V_L$ and $T_L$.
{}From this we conclude that smajoron and majorino decouple when they
are relativistic. Such hot relics can  cause the problem of
the excessive energy density if the particles are late-decaying.
In regards of inflation this problem should be examined in two ways.
The usual see-saw mechanism assumes a large scale $V_L \sim 10^{12}$
GeV for the lepton number breaking. It is bigger than the reheating
temperature $T_R$ which should be less than about $10^{10}$ GeV in order to
avoid the gravitino problem \cite{ekn}. In such a case ($V_L$ and $T_L > T_R$)
the relic density of majorino or smajoron can be washed away.
However, there can be a significant amount of regerated relics
after the reheating.  We will later consider the effects of regeneration.

It can also happen that smajoron and majorino decouple after inflation
($V_L$ or $T_L < T_R$).  For this case we note that the
crucial model-dependence of the cosmological problem resides,
especially, on the values of $x$ and $m_\psi$.
We begin with the discussion of this case. The ratio of the relic density
to the entropy density is then of order $10^{-3}$.
In the case that the smajoron decays into two majorons,
the decay-produced majorons
should be red-shifted away  sufficiently in order not to upset the standard
nucleosynthesis prediction. The majorons from smajoron decay should not
supply an energy density larger than three tenths of the density of one
neutrino species, which implies
\be
   \left( m_\sigma \over 10^2 \GeV \right)
   \left( \tau_\sigma \over 1 \sec \right)^{1/2} < 8.1 \times 10^{-5} \,.
\ee
{}From the  lifetime given in eq.~(\ref{ts}) one finds
\be \label{smajoron}
 V_L < 1.8 \times 10^{10} x
             \left(\frac{m_\sigma} {10^2{\rm GeV}}\right)^{1/2} \GeV \;.
\ee
Thinking of models with nonuniversal scalar masses
one may expect $x=O(1)$ so that it seems difficult to push $V_L$ higher
than $10^{10}$ GeV. Taking the universal boundary conditions for the soft terms
the value of $x$ depends on the Yukawa coupling $f$ between majoron
superfield and RHNs. For the two explicit models we have
mentioned here the situation can be summerized as

\be
 V_L < 4 \times 10^8 f^2 (1+A^2)
     \left(\frac{m_\sigma}{10^2\GeV} \right)^{1/2} \GeV \; .
 \label{f}
\ee
When the value of $f$ becomes extremely small, the decay channel of smajoron
into two majorons is suppressed and
those into two neutrinos or two sneutrinos become important.
This is the region of
$$ m_\nu > 7.5 f_4^2  \left( m_\sigma \over
                     10^2 \GeV \right)^{1/2} \eV \,,  $$
where the constraint in ref.~\cite{mz} is applicable.

A more severe constraint comes from the decay of
majorinos. Being R-parity odd, each majorino will eventually produce at least
one LSP (ordinary neutralino $\chi^0$) whose mass in generally exceeds the
GeV-range. Then those relics of secondary LSP cause an overclosure of
the universe.  To avoid this problem the majorino should decay before the
decoupling time of neutralinos which is about
$ 10^{-6} ( 20 \GeV/m_{\chi^0})^2$ sec.
If the majorino is heavier than the sneutrino the decay
$\psi \rightarrow \nu \tilde{\nu}$ produces a strong bound~:
\be \label{nb}
   V_L < 10^2 \left( m_\nu \over 10 \eV \right)^{1/2}
              \left( m_\psi \over 10^2 \GeV \right)^{1/2}
              \left( 20 \GeV \over m_{\chi^0} \right) \GeV   \,.
\ee
One can also find a bound of a similiar order of magnitude if the decay channel
$\psi \rightarrow \nu \nu \tilde{H^0}$ \cite{mz} is allowed.

Of course the majorino can be the LSP istself in which case it is a warm
dark matter candidate. Such a majorino
has to fulfill the overclosure bound $m_\psi< 2$ keV~\cite{rtw} which
translates to a restriction on $f$
\be
 f < 1.8\times10^{-3}\left(\frac{10^2\GeV}{m_{3/2}}\right)^{1/2}A^{-1/2}\; .
 \label{f_restr}
\ee
The model has now to be such that the tree level mass is below $\sim 2$ keV.
Assuming that $x$
comes entirely from RG effects eqs.~(\ref{f_restr}, \ref{f}) give
\be
 V_L< 1.3\times 10^{3}\left(\frac{m_\sigma}{10^2\GeV}\right)^{1/2}
 \left(\frac{10^2\GeV}{m_{3/2}}\right)\frac{1+A^2}{A}\GeV
 \label{ckn}
\ee
which is quite restrictive. It should, however, be stressed that this
bound is influenced by the magnitude of RG effects and is therefore model
dependent. Taking a model with a light majorino and nonuniversal soft masses it
would be completely invalidated and provided $f$ satisfies
eq.~(\ref{f_restr}) the strongest constraint on $V_L$ would be given by
eq.~(\ref{smajoron}).

\bigskip

Let us now consider the case that smajoron and majorino decouple before or
during inflation ($V_L$ and $T_L > T_R$). As already
mentioned our concern is the regeneration of relics.
In terms of the ratio of relic density to entropy density $Y_X=n_X/s$
the Boltzmann equation of the regenerated population for a particle $X$ becomes
\begin{equation}
\frac{dY_X}{dT} = -\frac{\langle\Sigma_Tv\rangle n_r^2}{sHT}
\end{equation}
where we neglected the decay terms.
We can explicitly integrate the Boltzmann
equation from the reheating temperature $T_R$ to some temperature $T$
with an initial condition $Y(T_R)=0$.
If $T_R\gg T$, we get
\begin{equation}
Y_X(T) \simeq \frac{\langle\Sigma_Tv\rangle n_r(T_R)^2}{H(T_R)s(T_R)}
       = 5.9\times10^{-6}\ \langle\Sigma_Tv\rangle M_{Pl} T_R \,.
\end{equation}
We see that $Y_X$ is determined by $\langle\Sigma_Tv\rangle$ and $T_R$.
For our case, the D=5 interaction in eq.~(\ref{d5}) is most important
for the regeneration and
$\langle\Sigma_Tv\rangle\simeq g^2 h^4/ 16^3 \pi^5 V_L^2$.
Then we obtain the regenerated population for smajoron or majorino~:
\be
 Y_{\sigma, \psi} \simeq 6.5 \times 10^{-11} f_1^2
                \left( m_\nu \over 10 \eV \right)^2
                 \left( T_R \over 10^{10} \GeV \right)\, .
\ee

We first consider the relevant constraints coming from the regeneration of
majorinos.  If the majorino is stable or its lifetime $\tau_\psi$ is longer
than the age of the universe $t_0$, the relation
\be \label{b1}
 Y_\psi  \left( m_\psi  \over 10^2\GeV \right)  < 3.55 \times 10^{-11}
\ee
has to be fulfilled in order not to overclose the universe.
This puts a bound on the neutrino mass
\be \label{bb1}
  m_\nu < 17  f_1^{-2}
                       \left( 10^{10} \GeV \over T_R \right)^{1/2}
                       \left( 20 \GeV \over m_\psi \right)^{1/2} \eV \,.
\ee
If the majorino is unstable and $\tau_\psi < t_0$, the overclosure
density bound is applied to the decay-produced neutralinos.  For this we
just replace $m_\psi$  by $m_{\chi^0}$ in the above equation.

There is another contstraint coming from the high-energy neutrinos when
the majorino (smajoron) is unstable.
The decay of smajoron and majorino generally produces high-energy neutrinos
which can be observed by experiments.  For the energy range of our interest,
the best bounds come from the experimental upper limit on upward-going muons
{}from the IMB detector~\cite{eglns}.  Here we quote the rough order
of estimation
\bea \label{b4}
  Y_\psi \left( m_\psi \over 10^2 \GeV \right)^2
            &<& 10^{-18} \left( t_0 \over \tau_\psi \right)^{4/3}
               \quad{\rm for}\quad \tau_\psi < t_0 \nonumber \\
          &<& 10^{-18} \left( \tau_\psi \over t_0 \right)
                \quad{\rm for}\quad \tau_\psi > t_0  \, .
\eea
Considering the decay of the majorino into a neutrino and
a sneutrino, we find for $\tau_\psi < t_0$
\be\label{bb4}
  V_L < 1.6 \times 10^{10} f_1^{-3/4} \left( m_\nu \over 10 \eV \right)^{1/4}
                    \left( 10^2 \GeV \over m_\psi \right)^{1/4}
                    \left( 10^{10} \GeV \over T_R \right)^{3/8} \GeV \, ,
\ee
and for $\tau_\psi > t_0$
\be  \label{bb5}
  V_L > 1.1\times 10^{17} f_1 \left( m_\nu \over 10 \eV \right)^2
                       \left( m_\psi \over 10^2 \GeV \right)^{3/2}
                       \left( T_R \over 10^{10} \GeV \right)^{1/2} \GeV \, .
\ee
Here the majorino lifetime is given by
\be
  \tau_\psi = 1.7 \times 10^{15} \left( V_L \over 10^{12} \GeV \right)^2
                            \left( 10 \eV \over m_\nu \right)^2
                            \left( 10^2 \GeV \over m_\psi \right) \sec \,.
\ee

The above results are summarized in the figures.
Fig.~1 shows the forbidden region in the $V_L-m_\nu$ space when the decay
$\psi \rightarrow \nu \tilde{\nu}$ is allowed.
The lines (ii) or (iii) are very sensitive to the value of $f$ and they provide
no constraint unless $f$ is larger than about $10^{-4}$.
As one expects, a rather strong constraint can be found  for larger
values of $f$ and $T_R$.  We note that $V_L$ is limited between
$T_R$ and the value in eq.~(\ref{bb4}) in order to have, e.g.,
a 40 eV neutrino as a dark matter candidate.
If the reheating temperature is pushed
up to $10^{10}$ GeV, the line (iv) goes down below the line  (a) and
we find a strong bound on the neutrino mass:
\be \label{gb}
  m_\nu < 18\, f_3^{-2/3} \left( 10^2 \GeV \over m_\psi \right)^{1/2}
                \left( 10^{10} \GeV \over T_R \right)^{1/6} \eV \,.
\ee

The size of the constrained region becomes smaller if the lifetime of the
majorino is fairly long or if it is stable.
When majorino is stable, only the lines (ii) and (iii)  are to be taken.
As an example for a
long-lived majorino which can decay into the LSP, let us take the decay
$\psi \rightarrow \nu \nu \tilde{H^0}$ with the Higgsino $\tilde{H^0}$
as the LSP. The lifetime can be found in ref.~\cite{mz} and together with
the constraint from the high-energy neutrinos it results in
the lower part of Fig.~2. The constrained region becomes smaller for
larger $T_R$ as the curve moves below the line (a).
For instance, there is no region forbidden if $T_R = 10^{10}$ GeV.

Regenerated smajorons play a less important role than majorinos.
With $x \sim 1$ one finds no constraints since then smajorons decay fast
enough. In a model with the RG-induced $x \sim 10^{-2} f^2$, the smajoron
decay into two neutrinos can be important for small values $f < 10^{-3}$,
for which the constraint from the high-energy neutrino
is the same as in  eqs.~(\ref{bb4}, \ref{bb5}).
The effect increases for smaller values of $f$ and larger values
of $T_R$. This behavior is shown in the upper curve in Fig.~2.

\bigskip

In conclusion we have discussed the cosmological constraints on SUSY majoron
models. The majoron superpartners decouple when relativistic and  we should
distinguish two cases, the decoupling temperature is lower or
higher than the reheating temperature after inflation.

In the first case we identified the mass of the majorino $m_\psi$ and the
quantity $x$ which determines the strength $x/V_L$ of the decay
$\sigma\rightarrow 2J$ as the crucial parameters influencing the
cosmological constraints. From the decay of the smajoron we obtained a
bound $V_L < x10^{10}$ GeV where in models with large RG effects or
nonuniversal scalar soft masses one can think of $x$ as being of $O(1)$.
Assuming standard soft scalar masses, however, in our examples $x$ was
determined by small RG effects leading to typical values $x\sim 10^{-2}f^2 $
where $f$ is the coupling of the majoron to the RHNs. Stronger bounds were
found from the majorino decay. If the majorino can decay into the LSP,
$V_L$ is constrained to be $< O(10^2$ GeV). However, there exist models with
a negligible tree level majorino mass and a radiative mass contribution
proportional to $f$. In such models the majorino can be the LSP and to avoid
overclosure $f$ has to fulfill $f<O(10^{-3})$. If at the same time $x$ is
generated by RG effects, a small $f$ surpresses the decay
$\sigma\rightarrow 2J$ which results in a typical bound $V_L<O(10^3$ GeV).
It should be stressed that this bound is avoidable~:
A model with a keV majorino and $x$ of order one would be fully consistent
up to $V_L\sim 10^{10}$ GeV.

For the second case we examined the effects of the regenerated relics.
The constrained region in the $V_L-m_\nu$ plane is strongly dependent on
the values of $f$ and the reheating temperature $T_R$ as well as the decay
channel of the majorino.
So  one needs more informations on the various parameters
to draw some definite conclusions.
Capitalizing the general pattern, we find a relatively strong constraints
if the majorino decays into a sneutrino and a neutrino is allowed.
In this case a higher value of $V_L$
together with a 10 eV neutrino can be allowed for a small value of the Yukawa
coupling $f< O(10^{-3})$.  If the above decay channel is not allowed,
we conclude that a higher value of $V_L$ can be easily achieved.
In either case the bounds on heavy neutrinos tend to be more restrictive.
We finish by making a remark on the  recent work \cite{msy} which showed that
the supersymetric singlet majoron model can explain both chaotic inflation and
baryogenesis.  In this scenario the inflaton is identified with the scalar
partner of a RHN. Then the primordial density perturbation
fixes the RHN mass at  $M \simeq 10^{13}$ GeV.
Furthermore the reheating temperature of order $10^{10}$ GeV could be obtained
with reasonable values of parameters.
Therefore this scenario should be such that the decoupling of majorino and
smajoron occurs before the reheating and should follow the constraints we
have investigated. For instance, assuming a Yukawa coupling $f$ of order
one, the above discussion  shows that the majorino should be lighter than
the sneutrino.

\bigskip

{\bf Acknowledgements.}
One of the authors (EJC) would like to thank G. Senjanovic and A. Yu.~Smirnov
for many helpful discussions and reading the manuscript.
This work was supported by
KOSEF through Center for Theoretical Physics, Seoul National University
and by the DFG-KOSEF-project 446 KOR-113/15/0 and the EC under contract
no.~SC1-CT91-0729.

\newpage

\section*{Figure captions}
\begin{itemize}
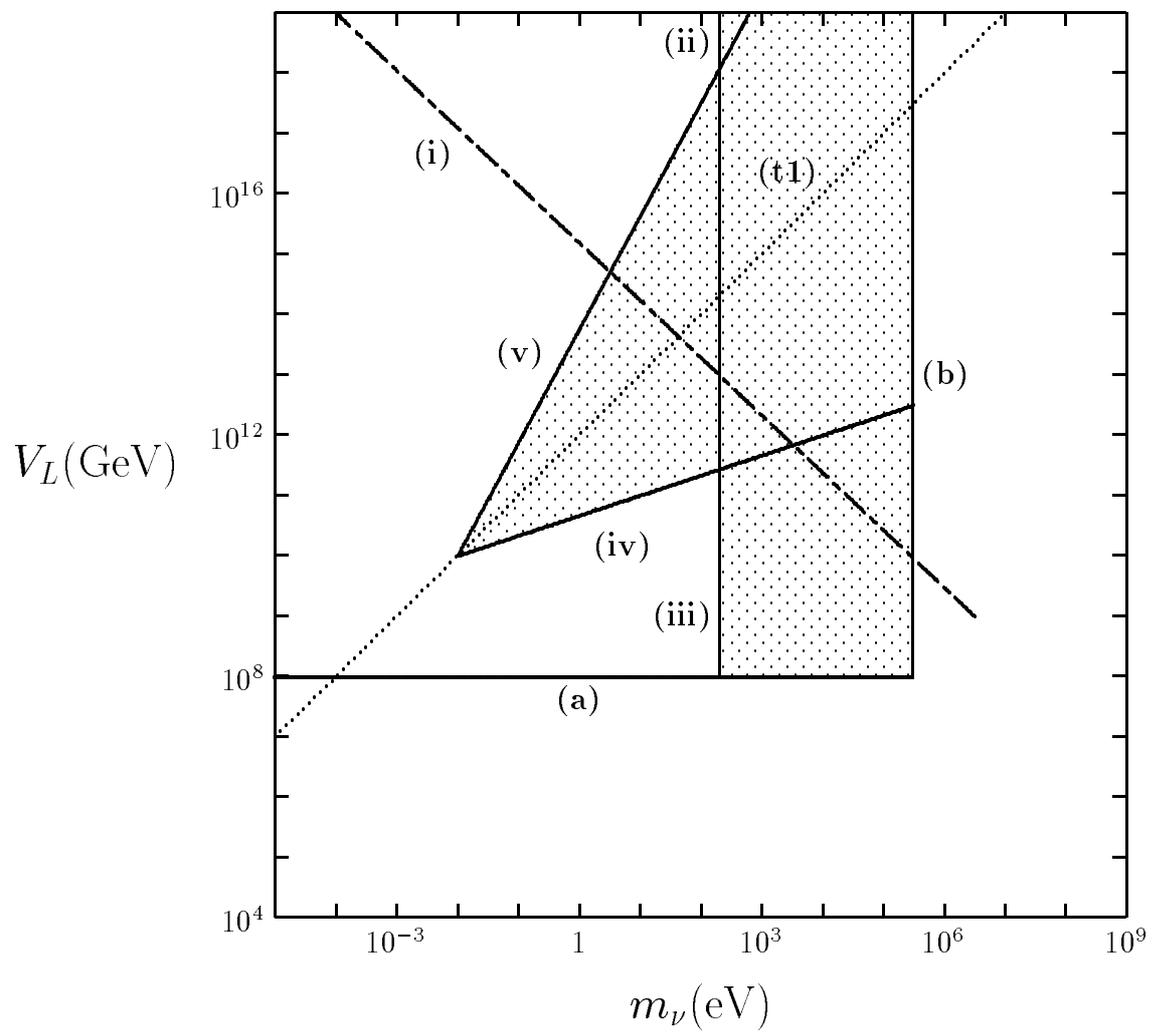
\item Fig.~1. For the decay $\psi \rightarrow \nu \tilde{\nu}$. The curves are
              normalized to $T_R = 10^{8}$ GeV and $f=0.1$.
       \begin{itemize}
       \item The lines (a) and (b) define the region where $V_L>T_R$ and
               $T_L > T_R$, respectively. The region below the line (t1)
               corresponds to $\tau(\psi \rightarrow \nu\tilde{\nu}) < t_0$.
       \item  Below the line (i) the yukawa coupling $h$ remains in
               perturbative  region $h< 1$, that is,
              $$ \left( V_L \over 10^{13} \GeV \right)
                    \left( m_\nu \over 10 \eV \right) < 6.1/f_1 \,. $$
       \item  The lines (ii) and (iii) arise from the overclosure bound for
              (quasi-) stable majorinos and decay-produced neutralinos,
               respectively. The region left to the lines is allowed.
               (cf.~eq.~(23) in the text.)
       \item  The lines (iv) and (v) show the constraints from
              the decay-produced  energetic neutrinos.
              The region below the line (iv) and above the line  (v) is
allowed.
               (cf. eqs.~(25, 26))
       \end{itemize}
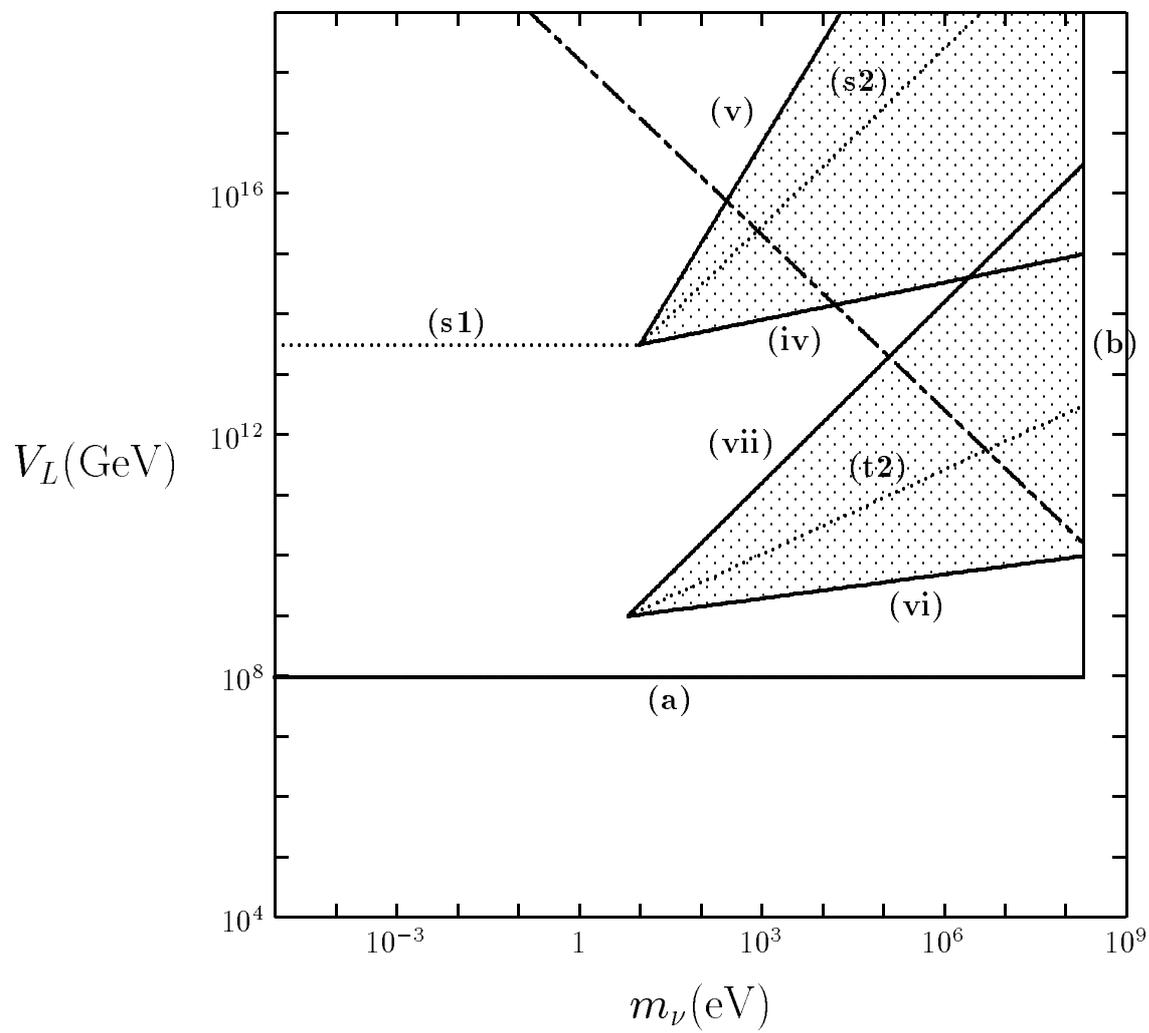
\item Fig.~2. For the decay $\psi \rightarrow \nu \nu \tilde{H}^0$ and
               $\sigma \rightarrow J J, \nu \nu$. The curves are
              normalized to $T_R = 10^{8}$ GeV and $f=10^{-4}$.
       \begin{itemize}
       \item The lines (a), (b) and (i)--(v) are as explained in Fig.~1.
           The lines (ii) and (iii) are now outside the experimental limit of
           $m_\nu$.
       \item The lines (s1, s2) represent $\tau(\sigma \rightarrow J J,\,
              \nu \nu) = t_0$. The line (t2) corresponds to
             $\tau(\psi \rightarrow \nu\nu\tilde{H}^0) = t_0$.
       \item  The lines  (vi) and (vii) show the constraints from
              the decay-produced energetic neutrinos as in eqs.~(25, 26).
              The region below the line (vi) is defined by
              $$  V_L < 1.2 \times 10^{8} f_3^{-7/8}
                         \left( m_\nu \over 10 \eV \right)^{1/8}
                    \left( m_\psi \over 10^2 \GeV \right)^{7/8}
                    \left( 10^{8} \GeV \over T_R \right)^{3/16} \GeV \,.$$
               and the region above the line (vii) by
               $$ V_L > 7.6 \times 10^{8} \left( m_\nu \over 10 \eV \right)
                       \left( m_\psi \over 10^2 \GeV \right)^{7/4}
                       \left( T_R \over 10^{8} \GeV \right)^{1/4} \GeV \,.$$
        \end{itemize}
\end{itemize}

\end{document}